# Harnessing Large Language Models for Mental Health: Opportunities, Challenges, and Ethical Considerations

Hari Mohan Pandey
Computing & Informatics, Bournemouth University, United Kingdom
hpandey@bournemouth.ac.uk/ profharimohanpandey@gmail.com

**Abstract:** Large Language Models (LLMs) are transforming mental health care by enhancing accessibility, personalization, and efficiency in therapeutic interventions. These AI-driven tools empower mental health professionals with real-time support, improved data integration, and the ability to encourage care-seeking behaviors, particularly in underserved communities. By harnessing LLMs, practitioners can deliver more empathetic, tailored, and effective support, addressing longstanding gaps in mental health service provision. However, their implementation comes with significant challenges and ethical concerns. Performance limitations, data privacy risks, biased outputs, and the potential for generating misleading information underscore the critical need for stringent ethical guidelines and robust evaluation mechanisms. The sensitive nature of mental health data further necessitates meticulous safeguards to protect patient rights and ensure equitable access to AI-driven care. Proponents argue that LLMs have the potential to democratize mental health resources, while critics warn of risks such as misuse and the diminishment of human connection in therapy. Achieving a balance between innovation and ethical responsibility is imperative. This paper examines the transformative potential of LLMs in mental health care, highlights the associated technical and ethical complexities, and advocates for a collaborative, multidisciplinary approach to ensure these advancements align with the goal of providing compassionate, equitable, and effective mental health support.

## 1. Introduction

Large Language Models (LLMs) are increasingly being integrated into mental health care, offering significant opportunities to enhance the accessibility, personalization, and efficiency of therapeutic interventions. These AI-driven tools can assist mental health professionals by providing real-time support, enhancing the quality of care through data integration, and promoting care-seeking behaviors among individuals in need. By leveraging LLMs, practitioners can improve emotional support delivery and create more empathetic treatment paradigms, potentially addressing critical gaps in mental health service provision, especially in underserved communities.[1][2][3] Despite their potential, the deployment of LLMs in mental health settings raises notable challenges and ethical concerns. Issues related to performance limitations, data privacy, and the risk of biased outputs can undermine the effectiveness of AI-driven interventions and exacerbate existing disparities in care. The sensitive nature of mental health data further complicates the use of these technologies, necessitating stringent ethical guidelines to safeguard patient information and ensure equitable access to care.[4][2][5] Moreover, the possibility of misleading outputs poses risks for vulnerable individuals, highlighting the importance of cautious implementation and ongoing evaluation.[2][4] Controversies surrounding the integration of LLMs into mental health care primarily focus on the balance between innovation and responsibility. Advocates argue that these models can democratize access to mental health resources, while critics caution against potential misuse and the erosion of the human element in therapy. Continuous monitoring and stakeholder engagement are essential to navigate these complexities, ensuring that the benefits of AI technologies do not compromise the quality of care or patient rights.[4][3][5] In conclusion, harnessing LLMs for mental health presents a double-edged sword; while they offer transformative possibilities for enhancing care delivery, they also introduce critical ethical, technical, and societal considerations that must be carefully managed. As this field continues to evolve, a collaborative approach among clinicians, researchers, and technology developers is crucial to align technological advancements with the overarching goal of providing compassionate and equitable mental health care.[1][2][5]

The paper is organized as follows: Section 2 explores the opportunities that Large Language Models (LLMs) present in healthcare, particularly in mental health and well-being; Section 3 examines the challenges associated with the use and implementation of AI-driven approaches and LLMs in health settings; Section 4 addresses the ethical concerns surrounding the integration of AI and LLMs into mental health care; Section 5 highlights their applications in therapeutic practices; Section 6 outlines key future directions that require focused attention for the effective use of AI and LLMs in healthcare; and finally, Section 7 presents the conclusions drawn from the study.

## 2. Opportunities

### 2.1 Support for Mental Health Professionals

LLMs can also serve as valuable tools for mental health professionals by providing real-time support during patient interactions. They can assist in generating evidence-based responses to patient queries and help clinicians identify and address gaps in patient data, thereby enhancing the quality of care delivered [1]. By reducing the cognitive load on practitioners, LLMs can contribute to more empathetic and effective mental health treatment paradigms [4].

### 2.2 Enhancing Care Provision

Large Language Models (LLMs) present significant opportunities for improving mental health care across various domains. They can facilitate care-seeking behaviors by providing accessible information and support to individuals in need, thereby reducing barriers to accessing mental health resources. By leveraging LLMs in community care settings, practitioners can enhance the provision of emotional support and guidance to patients, potentially improving outcomes and patient satisfaction [4].

### 2.3 Data Integration and Interoperability

LLMs can also contribute to strengthening data sharing protocols and improving interoperability between Electronic Health Record (EHR) systems. The implementation of standards such as the Fast Healthcare Interoperability Resources (FHIR) can promote better integration of patient data, ensuring that AI models are trained on comprehensive and representative datasets. This approach can mitigate issues related to fragmented care and enhance the continuity of patient data, ultimately leading to more effective AI-driven mental health interventions [2].

### 2.4 Addressing Social Determinants of Health

Incorporating external datasets that include individual and community-level Social Determinants of Health (SDoH) factors can enrich mental health care by providing valuable context for patient assessments. Collaborations with social service agencies can enhance data collection on SDoH, allowing healthcare providers to offer more personalized and equitable care. Government initiatives that incentivize the capturing of SDoH data can further support these efforts, leading to improved policymaking and healthcare access for marginalized populations [2].

## 2.5 Promoting Inclusivity and Diversity

Efforts to expand the availability of diverse and inclusive datasets are critical for the equitable application of AI in mental health. Programs like the NIH All of Us Research Program aim to create partnerships with a million diverse participants to advance precision medicine. This approach not only enriches the data used to train AI models but also ensures that the algorithms developed are more representative of the entire population, reducing biases that can arise from non-representative samples [3].

## 2.6 Development of User-Friendly Interfaces

To maximize the effectiveness of LLMs in mental health care, it is essential to develop user-friendly interfaces that accommodate various levels of digital literacy. By creating accessible platforms for patient-reported outcomes and encouraging consistent healthcare engagement, providers can improve data completeness and patient involvement in their care processes. This can lead to better outcomes and a more empowered patient population [2].

## 3. Challenges

## 3.1 Performance and Technical Limitations

The deployment of large language models (LLMs) in mental health settings faces significant technical and performance challenges. Issues such as model limitations, generalizability, and context-awareness hinder the reliability and efficacy of AI applications in complex real-world scenarios [2]. Variability in model performance often raises concerns regarding robustness and transparency, emphasizing the need for continuous innovation and rigorous evaluation [2].

## 3.2 Ethical Considerations and Bias

While LLMs hold the promise of improving mental health support through enhanced accessibility and personalization, they are often criticized for perpetuating biases against certain patient groups. This can lead to disparities in the effectiveness of AI-driven interventions, ultimately undermining the objective of equitable care [2] [5]. The intimate nature of mental health discussions also heightens the risk of misunderstandings and misuse, necessitating a careful examination of user expectations and anthropomorphism in AI systems [2].

## 3.3 Data Privacy and Security

Privacy concerns are paramount when utilizing LLMs in mental health applications due to the sensitive nature of the data involved. The risk of sensitive data exposure raises alarms about the need for robust data protection measures and ethical standards [2]. Ensuring safety and reliability in AI systems is critical to preventing harmful content generation and maintaining user trust in mental health services [2].

## 3.4 Misleading Outputs and Misinformation

A pressing challenge associated with LLMs is their potential to produce misleading or contextually uninformed outputs. In settings where access to qualified mental health professionals is limited, incorrect predictions or advice can lead to significant harm, particularly for marginalized individuals who may already face barriers to care [2] [4]. False positives can perpetuate stigma and diminish trust in mental health systems, while false negatives may leave individuals without necessary support [4].

## 3.5 Balancing Innovation and Responsibility

There is an ongoing debate regarding the balance between fostering innovation in LLM technology and ensuring responsible use within mental health contexts. Advocates for open-source models argue that transparency can enhance trust and accountability, while critics warn of the risks associated with potential misuse [2] [4]. Continuous monitoring and evaluation mechanisms are essential to adhere to ethical guidelines and navigate the complexities of AI in high-stakes mental health domains [4].

## 4. Ethical Considerations

The integration of large language models (LLMs) in mental health care brings forth a myriad of ethical considerations that require careful examination and governance. These considerations are essential to ensure that AI technologies enhance, rather than hinder, the quality of care provided to individuals experiencing mental health issues.

## 4.1 Informed Consent

Informed consent is a cornerstone of ethical medical practice, providing patients the right to make informed choices about their care. In the context of LLMs, patients must be fully aware of how these technologies will be utilized in their treatment, including the potential risks and benefits associated with their use.[6][7]. The necessity for transparent communication regarding the role of LLMs in mental health interventions cannot be overstated; individuals should be empowered to consent or decline AI-assisted treatment based on their understanding of the implications involved.[8][7].

### 4.2 Privacy and Data Security

Maintaining strict confidentiality is critical in mental health services. Ethical guidelines dictate that all client information must be safeguarded according to established legal and professional standards.[8]. Patients must understand how their personal data is collected, stored, and potentially used, particularly in the face of evolving AI technologies. Concerns around data misuse underscore the importance of ensuring that users have clarity about their rights to privacy and confidentiality, which should be upheld by robust data governance frameworks.[9][8].

### 4.3 Ethical Development and Implementation

The ethical development of AI systems for mental health care necessitates the involvement of multiple stakeholders, including clinicians, researchers, and technology developers. A collaborative approach is essential to establish frameworks that prioritize ethical considerations throughout the design and application processes of LLMs in healthcare. [9] [10]. Continuous ethical reviews and stakeholder engagement are advocated to foster responsible innovation in this domain, ensuring that technologies benefit patients without compromising their rights or well-being.[11][9].

### 4.4 Equity and Access Disparities

Another critical ethical issue relates to health equity and the distribution of mental health services. LLMs hold the potential to enhance access to care, but there exists a risk of exacerbating existing disparities if not implemented thoughtfully. Ethical frameworks must ensure that advancements in technology do not lead to unequal access or treatment disparities among different populations.[11][9].

### 4.5 Risk-Benefit Analysis

A comprehensive risk-benefit analysis is essential when implementing AI-driven solutions in mental health care. This includes assessing the reliability and effectiveness of LLMs and their potential impact on clinical outcomes. Ethical governance should focus on balancing the benefits of increased access and efficiency against the risks of reliance on automated systems that may lack transparency and accountability. [11] [6].

## 5. Applications in Therapeutic Practices

Large Language Models (LLMs) are emerging as transformative tools in the field of mental health therapy, offering a range of applications that can enhance both the effectiveness and accessibility of therapeutic interventions. Their potential lies particularly in augmenting cognitive behavioral therapy (CBT), which is a widely used treatment for common mental health disorders such as anxiety and depression. Given the increasing demand for therapists, LLMs could help bridge the gap between patient needs and available mental health support [12] [13].

### 5.1 Enhancing Predictive Analytics

One significant application of LLMs in mental health is their ability to improve predictive analytics related to patient outcomes. By leveraging machine learning techniques, LLMs can assist in predicting hospital readmission risks and disease progression, thereby augmenting decision support for clinicians. This predictive capability not only aids in proactive patient management but also facilitates better resource allocation within healthcare systems [4]. The integration of such predictive tools into clinical workflows has the potential to enhance the therapeutic quality through personalization and evidence-based adaptations tailored to individual patient needs [4][1].

### 5.1 Supporting Therapeutic Interventions

LLMs can assist in therapeutic interventions by providing personalized support to patients. They have been shown to reduce barriers for individuals who may feel stigmatized about seeking help face-to-face, particularly in sensitive contexts such as schizophrenia [14]. By allowing patients to engage with therapy through digital platforms, LLMs can enhance self-expression and improve the therapeutic process when integrated effectively into clinical practices [1][14]. Additionally, self-guided, web-based CBT programs have emerged as a practical solution to address the shortage of trained CBT therapists. These programs often involve minimal therapist input, assigning patients online modules that they can complete at their own pace. While this approach is scalable and cost-effective, it can lack personalization [12]. Here, LLMs have the potential to respond flexibly to individual circumstances, making therapeutic content more relevant and personalized for patients [12][15].

## 5.2 The Human Element in Therapy

Despite the promising applications of LLMs, it is crucial to recognize the importance of the human element in psychotherapy. Studies indicate that the effectiveness of therapy can depend significantly on the relationship between the therapist and the patient, emphasizing that human clinicians play a critical role in therapeutic outcomes [16]. While LLMs can augment therapy by assisting in peripheral tasks, the nuanced understanding and empathy that human therapists provide remain irreplaceable [16][13]. Therefore, a balanced approach that recognizes both the capabilities and limitations of LLMs is essential for their effective integration into mental healthcare [13].

## 5.3 Ethical Considerations and Patient Involvement

The ethical implications of using LLMs in therapy are profound, particularly concerning patient consent and the management of sensitive data. It is essential for medical staff to obtain explicit consent from patients regarding how LLMs will be used in their care, ensuring that patients are fully informed about the risks and limitations involved [4] [14]. Establishing trust and cooperation between patients and healthcare providers is critical in enhancing the overall quality of treatment, ultimately improving patient satisfaction and outcomes [4][14].

## 6. Future Directions for Using LLMs in Healthcare

To fully harness the transformative potential of Large Language Models (LLMs) in mental health care while addressing the associated challenges and ethical concerns, several future directions should be explored:

## 6.1 Enhancing Model Accuracy and Contextual Understanding

Future research should focus on improving the contextual understanding and emotional intelligence of LLMs to ensure accurate and empathetic responses. This involves fine-tuning models with diverse and representative mental health datasets while minimizing biases to ensure equitable and inclusive care delivery.

## 6.2 Developing Robust Ethical Frameworks

Establishing comprehensive ethical guidelines is critical to govern the deployment of LLMs in mental health care. These frameworks should prioritize patient data privacy, informed consent, and transparency while addressing potential risks like bias, misuse, and unintended consequences.

### 6.3 Integration with Multidisciplinary Teams

The collaboration between AI developers, mental health practitioners, ethicists, and policymakers should be strengthened to ensure that the design and deployment of LLMs align with clinical needs and ethical standards. Training mental health professionals to effectively use AI tools can also facilitate seamless integration into therapeutic practices.

### 6.4 Advancing Personalization in Care

Future efforts should emphasize the development of adaptive LLM systems that can tailor therapeutic interventions based on an individual's specific mental health conditions, preferences, and cultural contexts. This can enhance the relevance and effectiveness of AI-driven support.

### 6.5 Ensuring Accessibility and Equity

Research should address the digital divide by exploring ways to make LLM-driven mental health tools accessible to underserved and remote communities. Partnerships with non-profit organizations and public health systems can help democratize access to these technologies.

### 6.6 Continuous Monitoring and Evaluation

Implementing mechanisms for real-time monitoring, feedback, and evaluation of LLMs in mental health applications is essential to identify and mitigate potential harms. Regular audits of model performance, biases, and outputs will enhance accountability and reliability.

### 6.7 Exploring Hybrid Therapeutic Models

Future studies could investigate hybrid models that integrate LLMs with traditional therapeutic approaches, ensuring that AI complements rather than replaces human care. This would maintain the human element in therapy while leveraging the efficiency of AI-driven tools.

### 6.8 Regulatory and Policy Development

Policymakers should develop regulations tailored to the mental health sector, focusing on the ethical use of LLMs. Clear guidelines for liability, accountability, and quality assurance can foster trust and responsible adoption of these technologies.

By addressing these future directions, stakeholders can ensure that LLMs contribute meaningfully to mental health care, promoting compassionate, effective, and equitable support for individuals worldwide.

## 7. Conclusions

The integration of Large Language Models (LLMs) into mental health care holds immense potential to transform therapeutic practices by enhancing accessibility, personalization, and efficiency. These AI-driven tools offer significant advantages in supporting mental health professionals, addressing gaps in service provision, and fostering better care-seeking behaviors, particularly in underserved communities. However, their deployment also raises critical challenges and ethical concerns, such as performance limitations, data privacy issues, biases, and the risk of providing misleading or harmful outputs.

To ensure the effective and ethical use of LLMs, it is essential to implement stringent safeguards, establish robust evaluation frameworks, and develop comprehensive ethical guidelines. Additionally, continuous monitoring and active stakeholder engagement are necessary to address concerns surrounding the erosion of the human element in therapy and the potential misuse of AI technology. The balance between innovation and responsibility must remain a central focus as we move forward.

The future of LLMs in mental health care lies in a collaborative approach among clinicians, researchers, ethicists, and technology developers. By fostering such collaboration, we can maximize the benefits of LLMs while mitigating risks and ensuring that these technologies align with the goal of providing compassionate, equitable, and effective mental health care for all.

## References

[1] Hua, Yining, et al. "Large language models in mental health care: a scoping review." *arXiv preprint arXiv:2401.02984* (2024).

[2] De Choudhury, Munmun, Sachin R. Pendse, and Neha Kumar. "Benefits and harms of large language models in digital mental health." *arXiv preprint arXiv:2311.14693* (2023).

[3] Cross, James L., Michael A. Choma, and John A. Onofrey. "Bias in medical AI: Implications for clinical decision-making." *PLOS Digital Health* 3.11 (2024): e0000651.

[4] Nazer, Lama H., et al. "Bias in artificial intelligence algorithms and recommendations for mitigation." *PLOS Digital Health* 2.6 (2023): e0000278.

[5] Saeidnia, Hamid Reza, et al. "Ethical considerations in artificial intelligence interventions for mental health and well-being: Ensuring responsible implementation and impact." *Social Sciences* 13.7 (2024): 381.

[6] Warrier, Uma, Aparna Warrier, and Komal Khandelwal. "Ethical considerations in the use of artificial intelligence in mental health." *The Egyptian Journal of Neurology, Psychiatry and Neurosurgery* 59.1 (2023): 139.

[7] Lawrence, Hannah R., et al. "The opportunities and risks of large language models in mental health." *JMIR Mental Health* 11.1 (2024): e59479.

[8] Plante, T. "The ethics of AI applications for mental health care." *Markkula Center for Applied Ethics* (2023).

[9] Evidence-Based Mentoring. (n.d.). *The ethics of digital mental health applications*. Retrieved December 13, 2024, from https://www.evidencebasedmentoring.org/the-ethics-of-digital-mental-health-applications/

[10] Li, Anqi, et al. "Automatic evaluation for mental health counseling using llms." *arXiv preprint arXiv:2402.11958* (2024).

[11] Mirzaei, Tala, Leila Amini, and Pouyan Esmaeilzadeh. "Clinician voices on ethics of LLM integration in healthcare: a thematic analysis of ethical concerns and implications." *BMC Medical Informatics and Decision Making* 24.1 (2024): 250.

[12] Hodson, Nathan, and Simon Williamson. "Can large language models replace therapists? Evaluating performance at simple cognitive behavioral therapy tasks." *JMIR AI* 3.1 (2024): e52500.

[13] Forbes Technology Council. (2024, September 20). *AI can provide therapy but can't replace therapists so far: Here's why*. Forbes. Retrieved December 13, 2024, from https://www.forbes.com/councils/forbestechcouncil/2024/09/20/ai-can-provide-therapy-but-cant-replace-therapists-so-far-heres-why/

[14] Ma, Yingzhuo, et al. "Integrating large language models in mental health practice: a qualitative descriptive study based on expert interviews." *Frontiers in Public Health* 12 (2024): 1475867.

[15] Shen, Hao, et al. "Are Large Language Models Possible to Conduct Cognitive Behavioral Therapy?." *arXiv preprint arXiv:2407.17730* (2024).

[16] Fierce Healthcare. (n.d.). *How AI can shine a light into mental health interventions: Study*. Retrieved December 13, 2024, from https://www.fiercehealthcare.com/ai-and-machine-learning/ai-can-crack-open-black-box-effective-mental-health-counseling-scale-study